%% file: JohannesSivenManuscript.tex
\documentclass[preprint, authoryear]{elsarticle}
\usepackage{amsmath,amsfonts,amssymb}
\usepackage{graphicx}
\usepackage{epsf}
\usepackage{here}
\usepackage{graphicx}
\usepackage[american]{babel}

 \newtheorem{thm}{Theorem}

 \newtheorem{prop}[thm]{Proposition}
 \newdefinition{defn}[thm]{Definition}
 \newdefinition{rem}{Remark }

\begin{document}

\title{Value-at-Risk Computation by Fourier Inversion with Explicit Error
Bounds}
\author[saxo]{Johannes Vitalis Siven\corref{cor1}}
\ead{jvs@saxobank.com}

\author[saxo]{Jeffrey Todd Lins}
\ead{jtl@saxobank.com}

\author[saxo]{Anna Szymkowiak-Have}
\ead{aha@saxobank.com}

\cortext[cor1]{Corresponding author}

\address[saxo]{Saxo Bank A/S, Philip Heymans All\'e 15, DK-2900 Hellerup, Denmark}

\begin{abstract}
The value-at-risk of a delta-gamma approximated derivatives
portfolio can be computed by numerical integration of the
characteristic function. However, while the choice of parameters
in any numerical integration scheme is paramount, in practice  it
often relies on ad hoc procedures of trial and error. For normal
and multivariate $t$-distributed risk factors, we show how to
calculate the necessary parameters for one particular integration
scheme as a function of the data (the distribution of risk
factors, and delta and gamma) \emph{in order to satisfy a given
error tolerance}. This allows for implementation in a fully
automated risk management system. We also demonstrate in
simulations that the method is significantly faster than the Monte
Carlo method, for a given error tolerance.
\end{abstract} \maketitle

\section{Introduction}
Value-at-risk calculations of large derivatives portfolios are an
important part of risk management in many financial institutions.
The calculations are often performed by the Monte Carlo method and
can be very time consuming --- the main bottleneck is the
repricing of the portfolio for different simulated market
scenarios, in particular when the portfolio contains exotic
contracts that can only be priced with finite difference schemes
or even simulations. A common way to speed up computations is to
consider a delta-gamma approximation of the portfolio: the change
in the portfolio value over a fixed time horizon is approximated
by the quadratic function
$$
\Delta V = \delta ' \Delta S + \frac{1}{2} \Delta S' \Gamma \Delta
S,
$$
where $\Delta S$ is a vector of changes in the underlying risk
factors, and $\delta$ and $\Gamma$ are the usual first and second
order sensitivities. These are typically available ``for free''
since most often they are already computed internally for hedging
purposes. Under certain assumptions on the distribution of the
risk factors $\Delta S$, the characteristic function $\phi(u) =
\mathbb{E}[e^{iu\Delta V}]$ has a closed form expression. One can
write the cumulative distribution function as
\begin{equation}\label{eq:integral}
\mathbb{P}(\Delta V \leq x) = \frac{1}{2} - \int_{-\infty}^\infty
\frac{\phi(u)}{2\pi i u}e^{iux}du,
\end{equation}
and several authors have proposed to compute the value-at-risk by
evaluating this (or some related) integral numerically, thus
avoiding Monte Carlo simulations in the value-at-risk
computations, see \cite{Rouvinez} and \cite{DuffiePan}. In this
paper, we describe in detail how to perform this computation,
taking into account that the parameters in the numerical
integration scheme must be chosen as a function of the data (that
is, the distribution of the risk factors, and $\delta$ and
$\Gamma$).

Any numerical integration scheme involves choosing certain
parameters and for a particular case it is typically not hard to
find suitable parameters by trial and error. This might not be
good enough in an automated system: one would like the parameters
to be calculated as an internal part of the algorithm, preferably
with a guaranteed upper bound on the numerical error.

There is substantial literature on Fourier series methods for
computing tail probabilities, see the extensive reviews by
\cite{AbateWhitt} and \cite{FourierReview}, and the references
therein. Explicit error bounds are however rare: we use a result
from \cite{HughettBounds}, which allows for computing the
probability in Equation (\ref{eq:integral}) \emph{with a
guaranteed upper bound $\varepsilon$ on the error}, where
$\varepsilon> 0$ is specified by the user.

To apply the result, one must find certain constants that bound
the behavior of $\phi$ and the cumulative distribution function.
We show how to compute these constants as a function of the data
when the risk factors (i) are normally distributed, and (ii) have
a multivariate $t$-distribution. \cite{GlassermanVaR} derive a
closed form expression for the characteristic function in the
latter, heavy-tailed case. We also perform some simulation
experiments, where the Fourier method is benchmarked against the
Monte Carlo method. This is an important comparison to make, since
the Monte Carlo method is much faster for the delta-gamma
approximated portfolio than for the original portfolio (it avoids
the pricing bottleneck), and it is easy to understand and
implement.

The paper is organized as follows. In Section 2, we review the
delta-gamma approximation and recall the closed-form expressions
for the characteristic functions in the case of normally and
multivariate $t$-distributed risk factors. In Section 3, we state
the result from \cite{HughettBounds} on how to perform the
numerical integration with explicit error bounds, and discuss how
to choose the necessary parameters. Section 4 contains numerical
experiments, and Section 5 concludes.

\section{The delta-gamma approximation of a portfolio} \label{sec:deltagamma}

In this section we define the delta-gamma approximation of a
portfolio, and give the closed form expressions for the
characteristic function under two different assumptions on the
distribution of the risk factors.

 Let $S = (S_1,\ldots,S_p)'$ denote a vector
of $p$ risk factors to which a portfolio is exposed, and let
$\Delta S$ denote the change in $S$ from the current time 0 to the
end of the horizon $\Delta t$. Let $V(S,t)$ denote the value of
the portfolio at time $t$ and risk factors $S$. The delta-gamma
approximation to the change $V(S+\Delta S,\Delta t) - V(S,0)$ in
the portfolio value is given by
\begin{eqnarray*}
\Delta V:= \theta \Delta t + \delta ' \Delta S + \frac{1}{2}
\Delta S' \Gamma \Delta S,
\end{eqnarray*}
where
$$
\theta = \frac{\partial V}{\partial t}, \qquad \delta_i =
\frac{\partial V}{\partial S_i},\qquad \Gamma_{ij} =
\frac{\partial^2 V}{\partial S_i\partial S_j},\qquad i,j =
1,\ldots, p,
$$
and all partial derivatives are evaluated at $(S,0)$. The aim is
to compute VaR$_\gamma$, the level-$\gamma$ value-at-risk of
$\Delta V$, under certain assumptions for the distribution of
$\Delta S$. The level-$\gamma$ value-at-risk is defined as the
$\gamma$-quantile of the distribution of $\Delta V$:
$\mbox{VaR}_\gamma := P^{-1}(\gamma)$, where $P(x) :=
\mathbb{P}(\Delta V \leq x)$ is the cumulative distribution
function of $\Delta V$. Without loss of generality, we look only
at the case $\theta = 0$: if $\mathbb{P}(\Delta V - \theta \Delta
t \leq x) = \gamma$, then $\mathbb{P}(\Delta V \leq x +
\theta\Delta t) = \gamma$.

In the case of normally distributed risk factors $\Delta S$, the
characteristic function of $\Delta V$ is known on closed form.

\begin{prop} \label{Cf}
Assume that $\Delta S \sim \mathcal{N}(0,\Sigma)$ for some
positive definite matrix $\Sigma$. Let
$\lambda_1,\lambda_2,\ldots,\lambda_p$ be the eigenvalues of
$\Sigma \Gamma$, and let $\Lambda$ be the diagonal matrix with
these eigenvalues on the diagonal. There is a matrix $C$
satisfying $CC'= \Sigma$ and $C'\Gamma C =\Lambda$. Let $b =
C'\delta$. Then the characteristic function corresponding to $P$
is given by
$$
\phi(u) = \mathbb{E}[e^{iu\Delta V}]= e^{-\frac{1}{2}\sum_{j =
1}^pb_j^2\frac{u^2}{1 - i\lambda_j u}} \prod_{j =
1}^p(1-i\lambda_j u)^{-1/2}, \mbox{ for }u\in \mathbb{R}.
$$
The moment generating function is given by $\psi(u) = \phi(-iu)$
provided $u \lambda_j < 1$ for $j= 1,\ldots,p$.
\end{prop}

This result is certainly not new, but the proof is constructive
and useful for implementation.\footnote{The distribution of
quadratic forms of random variables is extensively studied by
\cite{MathaiProvost}.} We give a proof in the Appendix.

\cite{GlassermanVaR} relax the assumption of normal risk factors
and look at a heavy-tailed distribution for $\Delta S$. They
assume that $\Delta S$ has a multivariate $t$-distribution with
degrees-of-freedom parameter $\nu$: the density of $\Delta S$ is
given by
$$
f_{v,\Sigma}(x) =
\frac{\Gamma(\frac{1}{2}(p+\nu))}{(\nu\pi)^{p/2}\Gamma(\frac{1}{2}\nu)|\Sigma|^{1/2}}\left(1
+ \frac{1}{\nu}x'\Sigma^{-1}x \right)^{-\frac{1}{2}(p+\nu)},
\qquad \mbox{for }x\in \mathbb{R}^p,
$$
for some positive definite matrix $\Sigma$. Here, and only here,
$\Gamma(\cdot)$ denotes the gamma-function; elsewhere throughout
the paper, $\Gamma$ is the matrix of second order sensitivities in
the delta-gamma approximation of the portfolio. All the marginal
distributions of $\Delta S$ are $t$-distributions with
degrees-of-freedom parameter $\nu$. \cite{GlassermanVaR} show
that, given $x \in \mathbb{R}$, $P(x) = \mathbb{P}(\Delta V \leq
x) = F_x(0)$, where $F_x$ is another distribution function, and
derive a closed form expression for the characteristic function
corresponding to $F_x$.

\begin{thm}[Glasserman, Heidelberger and Shahabuddin] \label{tCf}
Denote the eigenvalues of $\Sigma \Gamma$ by
$\lambda_1,\lambda_2,\ldots,\lambda_p$ and let $\Lambda$  be the
diagonal matrix with these eigenvalues on the diagonal. There is a
matrix $C$ satisfying $CC' = \Sigma$ and $C'\Gamma C = \Lambda$.
Let $b = C'\delta$. Then $P(x) = F_x(0)$, where the characteristic
function corresponding to the distribution $F_x$ is given by
$$
\phi(u) = (1 -2\xi(u))^{-\nu/2} \prod_{j = 1}^p (1-i\lambda_j
u)^{-1/2},
$$
with
$$
\xi(u) = -\frac{iux}{\nu} - \frac{1}{2\nu}\sum_{j =
1}^pb_j^2\frac{u^2}{1 - i\lambda_j u}.
$$
The moment generating function is given by $\psi(u) = \phi(-iu)$,
provided $\lambda_j u < 1$ for $j= 1,\ldots,p$ and $\xi(-iu) <
\frac{1}{2}$.
\end{thm}

\begin{rem} \cite{GlassermanVaR} also consider a generalized
copula model, where the marginal distributions of $\Delta S$ are
$t$-distributions with possibly different degrees of freedom. The
generalization comes down to modifying $\delta$ and $\Gamma$ in a
straight forward way, and it applies here too.
\end{rem}

\section{Calculating VaR$_\gamma$ by Fourier inversion}
Let $F$ denote the cumulative distribution function for some
continuous distribution, and let $\phi$ denote the corresponding
characteristic function. Assume that $\phi$ is known on closed
form, but that $F$ is unknown. For a fixed $x\in \mathbb{R}$,
\cite[Theorem 10]{HughettBounds} tells us how to compute $F(x)$
with desired accuracy by numerical integration of $\phi$. In this
section we recall that result and discuss how to apply it in the
setting of the previous section for computing VaR$_\gamma$.

\begin{thm}[Hughett] \label{ErrBounds}
Suppose (i) that there exists constants $A$ and $\alpha >1$ such
that $F(-y) \leq A|y|^{-\alpha}$ and $1-F(y)\leq A|y|^{-\alpha}$
for all $y>0$, and (ii) that there exist constants $B$ and $\beta>
0$ such that $|\phi(u)|\leq B|u/2\pi|^{-\beta}$ for all $u\in
\mathbb{R}$. Then, for constants $0<l<2/3$, $T>0$ and $N>0$, the
distribution function $F(x)$ may be approximated by the truncated
Fourier series
$$
g(x) := \frac{1}{2} + 2\sum_{k =
1}^{N/2-1}\textrm{Re}\left(G[k]e^{i2\pi k x/T}\right),
$$
where Re$(\cdot)$ denote the real part, and
$$
G[k] := \frac{1-\cos(2\pi l k)}{i2\pi k}\phi(-2\pi k/T).
$$
The approximation error on the interval $|x|\leq l T/2$ is bounded
by
$$
|F(x) - g(x)| \leq \frac{2BT^{\beta}}{\pi}\zeta(\beta + 1,N/2) +
AT^{-\alpha}L_1(l,\alpha),
$$
where $\zeta$ denote the Hurwitz zeta function\footnote{The
Hurwitz zeta function is defined by $\zeta(z,a):= \sum_{k =
1}^\infty (k+a)^{-z}$. For computing it with desired accuracy, see
\cite[Lemma 4]{HughettBounds}.} and
$$
L_1(l,\alpha) := (l/2)^{-\alpha} + 2\zeta(\alpha,1 - \frac{1}{2}l)
+ \zeta(\alpha,1 + \frac{1}{2}l) +\zeta(\alpha,1 - \frac{3}{2}l).
$$
Furthermore, for any $D>0$ and $\varepsilon> 0$, choosing $l$, $T$
and $N$ such that
\begin{eqnarray}
0<l< \frac{2}{3} \ & \mbox{and}& \ l^\alpha
L_1(l,\alpha)\leq 2^{\alpha+1},\\
T\geq \frac{D}{l} \ &\mbox{and}& \ T \geq \frac{2}{l}\left(
\frac{3A}{\varepsilon}\right)^{1/\alpha},
\end{eqnarray}
and
$$
\qquad N \geq 2 + 2T\left( \frac{6 B}{\varepsilon\pi
\beta}\right)^{1/\beta}
$$
suffices to guarantee that $|F(x) - g(x)|\leq \varepsilon$ for
every $|x|\leq D/2$. It is always possible to choose $l$ to meet
the given conditions.
\end{thm}

\begin{rem}[Hughett] The conditions on
$l$, while complicated, involve only the value of $\alpha$. It is
thus feasible to precompute the smallest possible values of $l$,
for selected values of $\alpha$, using a numerical zero-finding
algorithm. Table 1 give the optimal values of $l$ for selected
values of $\alpha$.
\end{rem}

\begin{rem} \label{rem:computeVaR}
Using Theorem \ref{ErrBounds}, it is straight forward to compute
VaR$_\gamma$ both in the case of normally and multivariate
$t$-distributed risk factors. For normally distributed risk
factors, take $\phi$ from Proposition \ref{Cf} and look for a
solution to $g(x) = \gamma$ with any standard zero-finding method.
For multivariate $t$-distributed risk factors we similarly look
for a solution to $g_{x}(0)= \gamma$, where $g_x(0) = g(0)$ is
calculated from Theorem \ref{ErrBounds} with $\phi$ from
Proposition \ref{tCf}, where $\phi$ depends on $x$. In both cases
we end up with $x^*$ such that $|P(x^*) - \gamma|\leq \epsilon$.
\end{rem}

\input{Tab1}

Theorem \ref{ErrBounds} takes as input parameters $A$ and
$\alpha$, related to the unknown distribution function $F$, and
$B$ and $\beta$, related to the known characteristic function
$\phi$. The rest of this section discusses how to choose these
parameters in the setting of Section \ref{sec:deltagamma}.

\subsection{How to choose $A$ and $\alpha$}\label{chooseA} The computational
effort in using Theorem \ref{ErrBounds} is essentially
proportional to $(3A/\varepsilon)^{1/\alpha}$, so ideally $A$ and
$\alpha>1$ should minimize this quantity subject to
\begin{equation} \label{constr}
F(-y)\leq A|y|^{-\alpha} \mbox{ and } 1 - F(y)\leq A|y|^{-\alpha}
\mbox{ for all }y>0,
\end{equation}
where $F = P$ for normal risk factors, and $F = F_x$ for the
multivariate $t$-distribution. This problem might be difficult to
solve, since the constraints involve the unknown distribution
function $F$. It is however not important to find a truly optimal
solution: it suffices to make $(3A/\varepsilon)^{1/\alpha}$ as
small as possible with little computational effort. If the moment
generating function $\psi$ corresponding to $F$ is finite in a
neighborhood of zero,\footnote{If $\psi$ is finite in a
neighborhood of 0, then $\psi(u) = \phi(-iu)$, so this function is
also known on closed form.} one can use \emph{Chernoff's bounds},
see \cite{Chernoff}: $F(-y)\leq e^{-uy}\psi(-u)$ and $1 - F(y)\leq
e^{-uy}\psi(u)$ for all $y$ and all $u
> 0$. A sufficient condition for $A$ and $\alpha$
to satisfy the constraints (\ref{constr}) is thus that $A y
^{-\alpha}\geq \tilde{\psi}(u)e^{-uy}$, for all $y> 0$ and some $u
> 0$, where
$$
\tilde{\psi}(u):=\max\{\psi(u),\psi(-u)\}.
$$
Consider for a moment $u> 0$ as given, and assume that
$\tilde{\psi}(u)<\infty$. The following proposition gives the
solution to the optimization problem
\begin{equation}\label{eq:Asubopt}
\min_{A> 0,\alpha > 1} \left(
\frac{3A}{\varepsilon}\right)^{1/\alpha} \qquad \mbox{s.t. } A y
^{-\alpha}\geq \tilde{\psi}(u)e^{-uy},\mbox{ for all $y> 0$.}
\end{equation}

\begin{prop} \label{prop:Aopt}
The solution to the optimization problem (\ref{eq:Asubopt}) is
given by $\alpha(u) := \log (3\tilde{\psi}(u)/ \varepsilon)$ and
$A(u):= \tilde{\psi}(u)e^{-\alpha(u)}(\alpha(u)/u)^{\alpha(u)}$.
The minimal function value is $\log (3 \tilde{\psi}(u)/
\varepsilon )/u$.
\end{prop}
The proof is found in the Appendix. A tractable way of computing a
(suboptimal) minimizer of $(3A/\varepsilon)^{1/\alpha}$ that
satisfies the constraints (\ref{constr}) is to take $A = A(u^*)$
and $\alpha = \alpha(u^*)$, where $u^*$ solves
\begin{equation} \label{eq:Aopt}
\min_{u> 0} \log(3\tilde{\psi}(u)/\varepsilon)/u.
\end{equation}
This is a one dimensional optimization problem that can be solved
with standard methods.

\begin{rem}\label{rem:Aopt}
For normal risk factors, the optimization problem
(\ref{eq:Asubopt}) needs to be solved only once. The resulting $A$
and $\alpha$ can be used to compute $P(x)$ for any $x\in
\mathbb{R}$. For the multivariate $t$-distribution, $P(x) =
F_x(0)$, so the optimization problem needs to be solved once for
each $x$, each time using the moment generating function
corresponding to $F_x$.
\end{rem}

\subsection{How to choose $B$ and $\beta$}\label{chooseB} The computational
effort in using Theorem \ref{ErrBounds} is essentially
proportional to $(6B/\pi\varepsilon \beta)^{1/\beta}$, so for fast
computations one would like to choose $B$ and $\beta$ that
minimize this quantity, subject to $|\phi(u)|\leq
B|u/2\pi|^{-\beta}$ for all $u$. In the general case, solving this
problem comes down to analyzing the characteristic function. For
the two special cases of normally and multivariate $t$-distributed
risk factors, we have the following bounds. The proof is found in
the Appendix.

\begin{prop} \label{CfBounds}
Let $\lambda_1,\ldots,\lambda_p$ be the eigenvalues of $\Sigma
\Gamma$, and let $I$ denote any subset of $\{1,\ldots,p\}$. Let
$\beta_I := |I|/2$ and $B_I := (2\pi)^{-\beta_I}\prod_{j\in
I}|\lambda_j|^{-1/2}$. If $\phi$ is given either by Proposition
\ref{Cf} or Proposition \ref{tCf}, then $|\phi(u)| \leq B_I
|u/2\pi|^{-\beta_I}$.
\end{prop}

A tractable strategy for choosing $B$ and $\beta$ is to solve
\begin{equation}\label{Iopt}
\min_I
\left(\frac{6B_I}{\pi\varepsilon\beta_I}\right)^{1/\beta_I},
\end{equation}
where $\beta_I$ and $B_I$ are defined in Proposition
\ref{CfBounds}, and the minimum is taken over all the subsets of
$\{1,\ldots,p\}$.

\begin{prop}\label{prop:Iopt}
Let $\{j_1,\ldots,j_p\}$ be an ordering of $\{1,\ldots,p\}$, such
that $|\lambda_{j_1}|\geq |\lambda_{j_2}|\geq \cdots \geq
|\lambda_{j_p}|$. Denote $I_k := \{j_1,\ldots,j_k\}$ for $k =
1,\ldots,p$, and let $I$ denote a solution to the optimization
problem (\ref{Iopt}). Then $I = I_k$ for some $k \in
\{1,\ldots,p\}$.
\end{prop}

\begin{rem} \label{rem:Bopt}
By Proposition \ref{prop:Iopt}, the optimal subset $I$ (and thus
$\beta_{I}$ and $B_{I}$) can be found by computing
$(6B_{I_k}/\pi\varepsilon\beta_{I_k})^{1/\beta_{I_k}}$ for $k =
1,\ldots,p$ and choosing the minimum. Note that this needs to be
done only once, even in the case of multivariate $t$-distributed
risk factors --- although $\phi$ depends on $x$, the optimal
parameters $B_I$ and $\beta_I$ do not.
\end{rem}

\section{A large scale numerical experiment}
In this section the value-at-risk for a large, simulated options
portfolio is calculated with two different methods: (i) the
Fourier method from the previous section, and (ii) the Monte Carlo
method. The results are then compared with respect to speed and
accuracy.

\subsection{Simulating a large options portfolio} We simulate a
portfolio that models a real portfolio held by a market maker in
vanilla options: it contains $10^4$ vanilla options written on 30
different underlyings. Each option is long or short with
probability $\frac{1}{2}$, and a call or a put option with
probability $\frac{1}{2}$. The maturity of each option is drawn
from a uniform distribution on $[10/252,1]$ (10 days-1 year), and
the moneyness for each option is drawn from a truncated normal
distribution with mean 1 and standard deviation 0.1, bounded below
by 0.5 and above by 1.5 --- the options are between 50\%
out-of-the-money and 50\% in-the-money. The returns from the
underlyings are drawn (i) from a normal distribution or (ii) from
a multivariate $t$-distribution with $\nu = 5$ degrees of freedom.
In both cases, the covariance matrix is $\Delta t\Sigma$, where
$\Sigma$ is a $30\times 30$ diagonal matrix with each diagonal
element drawn from a uniform distribution on $[0.1^2,0.3^2]$ and
$\Delta t := 10/252$ (so we are looking at the 10 day
value-at-risk). The nominal value for each option is drawn from a
uniform distribution on $[10^4,10^5]$.

Having simulated the portfolio, we compute the deltas and gammas
for each option. Finally, the deltas and gammas are summed up for
each underlying, to give $\delta$ and $\Gamma$.

\subsection{Performance of the Fourier method} Given $\Sigma$,
$\delta$ and $\Gamma$ from the simulation, we compute
$\lambda_1,\ldots,\lambda_{30}$ and $b_1,\ldots,b_{30}$ in
Theorems \ref{Cf} and \ref{tCf}. As seen in the proof of
Proposition \ref{Cf}, this is done by computing the Cholesky
factorization of $\Sigma$, and diagonalizing the matrix $\Sigma
\Gamma$. This gives us the complete closed-form expression for the
characteristic function and the moment generating function. Let
$\varepsilon > 0$ be given, and compute the necessary parameters,
$A,\alpha, B$ and $\beta$, as described in Remarks \ref{rem:Aopt}
and \ref{rem:Bopt}. The level-$\gamma$ value-at-risk is then
computed as described in Remark \ref{rem:computeVaR}. The
algorithm gives a number $x^*$, which is close to VaR$_\gamma =
P^{-1}(\gamma)$ in the sense that $|P(x^*) - \gamma|\leq
\varepsilon$.

The computational effort is proportional to the integer $N$, so we
report this together with the total computational
time\footnote{The computations were performed in the software {\tt
R} on a standard workstation.} for various choices of
$\varepsilon$ in Table 2 (normal risk factors) and Table 3
(multivariate $t$-distributed risk factors).

\input{Tab2}

\input{Tab3}

\subsection{Performance of the Monte Carlo method} To compute an
estimate $x^*$ of VaR$_\gamma = P^{-1}(\gamma)$ by the Monte Carlo
method, simulate $M$ samples from the
$\mathcal{N}(0,\Sigma)$-distribution (or the multivariate
$t$-distribution), compute the corresponding values of $\Delta V$
and order these to get $\Delta V_ {(1)}\leq \cdots \leq \Delta
V_{(M)}$. Then $x^* := \Delta V_{(\lceil M\gamma \rceil)}$, where
$\lceil M\gamma \rceil$ is the smallest integer greater than or
equal to $M\gamma$. The computed $x^*$ corresponds to some
confidence level $\gamma^* := P(x^*)$, hopefully not too far from
$\gamma$. To study how large sample size $M$ we need to reach a
desired level of accuracy we look at the number $\varepsilon$ such
that $\mathbb{P}(|\gamma - \gamma^*|\leq \varepsilon) = 1-\rho$
for some small positive number $\rho$.

For a moderate sample size $M$, the corresponding $\varepsilon$
can be estimated experimentally in the following way: run the
Monte Carlo $R$ times, order the resulting estimates
$x^*_{(1)}\leq\cdots \leq x^*_{(R)}$ and compute the corresponding
confidence levels $\gamma_{(1)}^*\leq\cdots\leq\gamma^*_{(R)}$ by
evaluating $\gamma_{(k)}^*:= P(x^*_{(k)})$, for $k = 1,\ldots,R$,
using the Fourier method with an extremely low error. Then take
$$\varepsilon:=
\frac{1}{2}\left(\gamma^*_{(\lceil(1-\rho/2)R\rceil)} -
\gamma^*_{(\lfloor\rho R/2\rfloor)}\right),$$
where $\lfloor\rho
R/2\rfloor$ denotes the largest integer smaller than or equal to
$\rho R/2$.

For a large sample size $M$, we can compute $\varepsilon$ from the
approximate distribution of $\gamma^*$. Since $x^*$ is the $\lceil
M\gamma \rceil$th order statistic, it is asymptotically (as
$M\rightarrow \infty$) normally distributed\footnote{This is true
if $\Delta V$ has a continuous non-zero density at
$P^{-1}(\gamma)$, see \cite{orderStat}.} 
with mean VaR$_\gamma$ and variance
$\gamma(1-\gamma)/(MP'($VaR$_\gamma)^2)$. If $P$ is approximately
linear in a neighborhood of VaR$_\gamma$, then $\gamma^* = P(x^*)$
is approximately normally distributed with mean $\gamma$ and
variance $\gamma(1-\gamma)/M$ by Gauss' approximation formulas.
So, $\varepsilon \approx
\Phi^{-1}(1-\rho/2)\sqrt{\gamma(1-\gamma)/M}$, where $\Phi(\cdot)$
is the cumulative distribution function of the standard normal
distribution.

\begin{figure}
\begin{center}
\includegraphics[width=10cm]{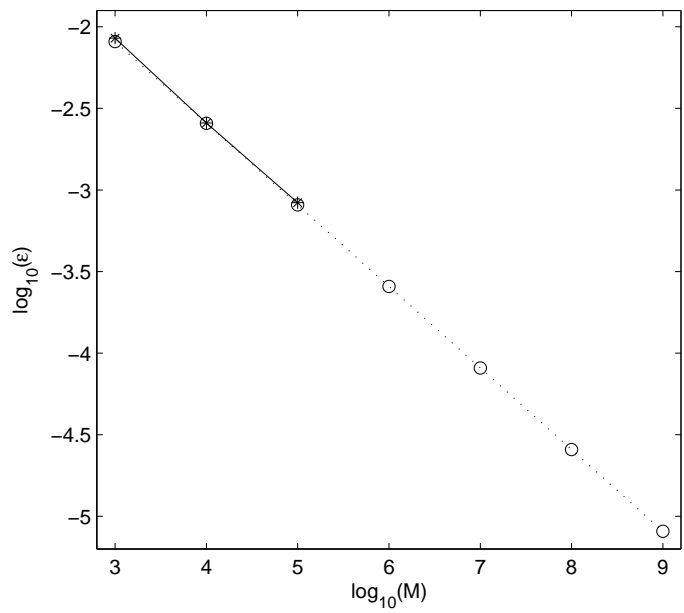}
\end{center}
\caption{The graph shows $\varepsilon$ as a function of the Monte
Carlo sample size $M$, where $\varepsilon$ is computed
experimentally (stars) and from the approximative theoretical
distribution of $\gamma^*$ (circles).} \label{fig:MC}
\end{figure}

Figure \ref{fig:MC} show $\varepsilon$ estimated experimentally
for $M = 10^3,10^4,10^5$, with $R = 500$ and $\rho = 0.01$. It
also shows $\varepsilon$ computed theoretically from the
approximate distribution of $\gamma^*$, for $M =
10^3,10^4,\ldots,10^9$ --- the two methods for computing
$\varepsilon$ agree nicely for $M \geq 10^3$. If $\varepsilon$ is
small enough, we can thus use the theoretical approximate
distribution of $\gamma^*$ to compute the sample size $M$ needed
to make $\mathbb{P}(|\gamma-\gamma^*|\leq \varepsilon) \approx 1-
\rho$ for a given $\varepsilon$ --- see Table 3 for $\rho = 0.01$
and various choices of $\varepsilon$. The total computing time for
$M = 6.5\cdot 10^4$ is 1.7 s for normally distributed risk
factors, and 2.4 s for multivariate $t$-distributed risk factors.
It is moderately more time consuming to simulate from the
multivariate $t$-distribution than from the normal distribution,
see \cite{GlassermanVaR}. 

\input{Tab4}

\subsection{Comparison of the Fourier method and the Monte Carlo
method} Comparing the results from the Fourier method and the
Monte Carlo method, we see that the Fourier method is superior
both with respect to speed and accuracy. For $\varepsilon =
10^{-3}$ and normally distributed risk factors, the Fourier method
takes 0.12 seconds to compute $x^*$ such that $|\gamma - P(x^*)|
\leq \varepsilon$, while the Monte Carlo method takes 1.76 s to
compute $x^*$ such that $\mathbb{P}(|\gamma - P(x^*)|\leq
\varepsilon)\geq 0.99$. As $\varepsilon \rightarrow 0$, the Monte
Carlo method becomes impractical: a 10 times smaller $\varepsilon$
gives 100 times longer computing time. The Fourier method however
scales nicely: for $\varepsilon = 10^{-6}$ the whole computation
still takes only about 0.3 s. The picture is similar for
multivariate $t$-distributed risk factors: the Fourier method is
significantly faster than the Monte Carlo method, especially for
very small $\varepsilon$.

\section{Conclusion}
We have given a complete description of how to compute the
value-at-risk of a delta-gamma approximated portfolio by numerical
integration of the characteristic function, in the case of
normally or multivariate $t$-distributed risk factors. In
particular, we showed how to calculate the necessary parameters in
the numerical integration scheme as a function of the data (the
distribution of the risk factors, and $\delta$ and $\Gamma$), in
order to satisfy a given error tolerance. Thus we avoid ad hoc
choices of parameters, and allow for using the method in a fully
automated risk management system. Our numerical experiments
illustrate that the method is significantly faster than the Monte
Carlo method --- we conclude that the Fourier method is a highly
competitive alternative to the Monte Carlo method for computing
the value-at-risk of delta-gamma approximated portfolios.
\bibliographystyle{agsm}
\bibliography{BibJohannesSiven}

\section*{Appendix}

\begin{bf}Proof of Proposition \ref{Cf}.\end{bf}
Let $L$ be the Cholesky factor of $\Sigma$, so $L L' = \Sigma $
and $L^{-1} \Delta S$ is a vector of independent, standard
normals. Since $\Sigma$ is a symmetric matrix, $L'\Gamma L$ has
real eigenvalues, and these eigenvalues are the same as those of
$LL'\Gamma = \Sigma \Gamma$, namely $\lambda_1,\ldots,\lambda_p$.
Moreover, $L'\Gamma L = O\Lambda O'$, where $O$ is an orthogonal
matrix whose columns are eigenvectors of $L' \Gamma L$. Since
$O'L'\Gamma L O = \Lambda$ and $(LO)(LO)' = L L' =\Sigma$, setting
$C = LO$ produces the required matrix.

Introduce the identity matrix $I = LOO'L^{-1} = (L^{-1})'OO'L'$ in
the expression for $\Delta V$:
\begin{eqnarray*}
\Delta V &=& \delta'\Delta S + \frac{1}{2}\Delta S'\Gamma \Delta
S\\
 &=&\delta'LOO'L^{-1}\Delta S +
\frac{1}{2}\Delta S'(L^{-1})'OO'L'\Gamma LOO'L^{-1}\Delta S\\
&=&(O'L'\delta)'(O'L^{-1}\Delta S) + \frac{1}{2}(O'L^{-1}\Delta
S)'(O'L'\Gamma LO)(O'L^{-1}\Delta S).
\end{eqnarray*}
Write $\tilde{\Delta S} =(\tilde{\Delta S}_1,\ldots,\tilde{\Delta
S}_p)' : = O'L^{-1}\Delta S$ and recall that $b = C'\delta =
O'L'\delta$. Then
\begin{eqnarray*}
\Delta V &=& b'\tilde{\Delta S} + \frac{1}{2}\tilde{\Delta
S}'\Lambda \tilde{\Delta S} = \sum_{j = 1}^pb_j\tilde{\Delta S}_j
+ \frac{1}{2}\sum_{j = 1}^p\lambda_j\tilde{\Delta S}_j^2,
\end{eqnarray*}
where $\tilde{\Delta S}_1,\ldots,\tilde{\Delta S}_p$ are
independent, $\mathcal{N}(0,1)$-distributed random variables. Let
$J := \{j\in \{1,\ldots,p\};\ \lambda_j\neq 0\}$ and write
\begin{eqnarray*} \Delta V &=& \sum_{j\not\in J}b_j\tilde{\Delta S}_j +
\sum_{j\in J}\left( b_j\tilde{\Delta S}_j +
\frac{1}{2}\lambda_{j}\tilde{\Delta S}_j^2\right)\\
&=& \sum_{j\not\in J}b_j\tilde{\Delta S}_j-\sum_{j\in
J}\frac{b_j^2}{2\lambda_{j}}  + \sum_{j\in
J}\frac{1}{2}\lambda_{j}\left( \tilde{\Delta S}_j +
\frac{b_j}{\lambda_{j}}\right)^2,
\end{eqnarray*}
where the second equality arises from completing the squares.
Define $Q_0 := \sum_{j\not\in J}b_j\tilde{\Delta S}_j -
1/2\sum_{j\in J}b_j^2/\lambda_{j}$, and $Q_j := (\tilde{\Delta
S}_j + b_j/\lambda_{j})^2$, for $j\in J$. Then $\Delta V = Q_0 +
1/2\sum_{j\in J}\lambda_j Q_j$ is a linear combination of
independent random variables --- $Q_0$ and $Q_j$, $j\in J$, are
independent since $\tilde{\Delta S}_1,\ldots,\tilde{\Delta S}_p$
are independent, so the characteristic function of $\Delta V$ is
given by $$\phi(u) = \mathbb{E}[e^{iu\Delta V}] =
\mathbb{E}[e^{iuQ_0}]\prod_{j\in
J}\mathbb{E}[e^{i(\lambda_j/2)uQ_j}].$$
The distribution of $Q_0$
is normal with mean $-1/2\sum_{j\in J}b_j^2/\lambda_{j}$ and
variance $\sum_{j\not\in J}b_j^2$, and $Q_j$ has a non-central
$\chi^2$-distribution\footnote{The characteristic function of a
$\chi^2$-distributed random variable $Z$ with $k$ degrees of
freedom and non-centrality parameter $\alpha$ is given by
$\mathbb{E}[e^{iuZ}]= e^{\alpha\frac{iu}{1-i2u}}(1-i2u)^{-k/2}$.}
with 1 degree of freedom and non-centrality parameter
$b_j^2/\lambda_j^2$, for $j\in J$. If follows that
\begin{eqnarray*}
\phi(u) &=& \exp\left(-i\frac{u}{2}\sum_{j\in
J}\frac{b_j^2}{\lambda_{j}} - \frac{u^2}{2}\sum_{j \not \in J}
b_j^2\right)
\prod_{j\in J} \exp\left(\frac{b_j^2}{\lambda_{j}^2}\frac{i\frac{\lambda_j}{2}u}{1-i\lambda_ju}\right)\frac{1}{\sqrt{1-i\lambda_j u}}\\
&=& e^{-\frac{1}{2}\sum_{j = 1}^pb_j^2\frac{u^2}{1 - i\lambda_j
u}} \prod_{j\in J}\frac{1}{\sqrt{1-i\lambda_j u}}.
\end{eqnarray*}

\begin{flushright} $\Box$ \end{flushright}

\begin{bf} Proof of Proposition \ref{prop:Aopt}.\end{bf}
Assume that $\tilde{\psi}(u)<\infty$ and consider for a moment
$\alpha > 1$ as given. Then the constraint will be satisfied if
$A$ is taken large enough. For the smallest possible $A$, the
graph of the function $f(y) = Ay^{-\alpha}$ will not intersect the
graph of $h(y) = \tilde{\psi}(u) e^{-uy}$, but be tangent to it at
some point $y^*$. This is expressed in the following system,
$$
\left\{
\begin{array}{rcl}
f(y^*) &=& h(y^*),\\
f'(y^*) &=& h'(y^*),\\
f''(y^*) &\geq& h''(y^*),
\end{array}\right.
$$
which is solved by $y^* = \alpha/u$ and $A(\alpha,u) =
\tilde{\psi}(u)e^{-\alpha}(\alpha/u)^{\alpha}$. Plugging the
expression for $A(\alpha,u)$ into the minimization problem
(\ref{eq:Asubopt}) reduces it to
$$
\min_{\alpha >
1}\frac{\alpha}{u}\left(\frac{3\tilde{\psi}(u)}{\varepsilon}\right)^{1/\alpha}.
$$
This target function has the unique minimum $\alpha(u) :=
\log(3\tilde{\psi}(u)/\varepsilon)$. The result follows, since
$(3A(\alpha(u),u)/\varepsilon)^{1/\alpha(u)} = \frac{1}{u}\log (3
\tilde{\psi}(u)/ \varepsilon)$.
\begin{flushright} $\Box$ \end{flushright}

\begin{bf}Proof of Proposition \ref{CfBounds}.\end{bf}
If $\phi$ is given by Proposition \ref{Cf}, then $\phi(u) =
\phi_1(u)\phi_2(u)$, with
$\phi_1(u):=\exp\left(-\frac{1}{2}\sum_{j = 1}^pb_j^2\frac{u^2}{1
- i\lambda_j u}\right)$ and  $\phi_2(u):=\prod_{j = 1}^p (1-i
\lambda_j u)^{-1/2}$. If $\phi$ is given by Proposition \ref{tCf},
then $\phi(u) = \phi_3(u)\phi_2(u)$ with $\phi_3(u) := (1 -
2\xi(u))^{-\nu/2}$. Since $u^2/(1-i\lambda_j u) =
\frac{u^2}{1+\lambda_j^2u^2} +
i\frac{u^3\lambda_j}{1+\lambda_j^2u^2}$, it follows that
$$
|\phi_1(u)| =e^{-\frac{1}{2}\sum_{j=1}^p
b_j^2\frac{u^2}{1+u^2\lambda_j^2}} \leq 1.$$ It also follows that
$$
|1 - 2\xi(u)|= \left|1 + \frac{1}{\nu}\sum_{j =
1}^pb_j^2\frac{u^2}{1 +\lambda_j^2 u^2} +
i\textrm{Im}(-2\xi(u))\right|\geq 1,
$$
so $|\phi_3(u)| = |1 - 2\xi(u)|^{-\nu / 2} \leq 1$.

 To bound $|\phi_2(u)| \geq \lambda_ju|$, first note that
$|1-iu\lambda_j| = \sqrt{1 + \lambda_j^2u^2}\geq
\max\{1,|\lambda_ju|\}$, so $|1-i\lambda_j u|^{-1/2}\leq
\min\{1,|\lambda_ju|^{-1/2}\}$. This implies that
$$
|\phi_2(u)| = \prod_{j  = 1}^p |1-i\lambda_j u|^{-1/2} \leq
\prod_{j \in I}|\lambda_j|^{-1/2}|u|^{-1/2} = |u|^{-|I|/2}\prod_{j
\in I}|\lambda_j|^{-1/2}
$$
for any subset $I\subset \{1,\ldots,p\}$. It follows that
$|\phi(u)|\leq B|u/2\pi|^{-\beta}$ with $\beta = |I|/2$ and $B =
(2\pi)^{-\beta}\prod_{j\in I}|\lambda_j|^{-1/2}$ both for $\phi =
\phi_1\phi_2$ and $\phi = \phi_3 \phi_2$.
\begin{flushright} $\Box$ \end{flushright}

\begin{bf}Proof of Proposition \ref{prop:Iopt}.\end{bf}
Assume the contrary to the Proposition: $I \neq I_k$ for $k =
1,\ldots,p$. Then there exists $m,m' \in \{1,\ldots,p\}$, $m< m'$,
such that $j_m\not \in I$ and $j_{m'}\in I$. Let $I' :=
(I\setminus\{j_{m'}\})\cup\{j_m\}$. Then $\beta_I = \beta_{I'}$
and, unless $|\lambda_{j_m}| = |\lambda_{j_{m'}}|$, $B_{I'}<B_I$
--- this contradicts the optimality of $I$. The degenerate case
$|\lambda_{j_m}| =|\lambda_{j_{m'}}|$ only arises from the trivial
ambiguity in ordering numbers of the same size. To rule it out, we
adopt the following convention: if $s$ indices
$j_{t+1},\ldots,j_{t+s}$ correspond to
$|\lambda_{j_{t+1}}|=\cdots=|\lambda_{j_{t+s}}|$ and if $I$
contains $k<s$ of these indices, then it contains the $k$ first
ones: $j_{t+1},\ldots,j_{t+k}$.
\begin{flushright} $\Box$ \end{flushright}
\end{document}

%% file: Tab1.tex
\begin{table}[t]
\begin{center}
  {\scriptsize
  \begin{tabular}{cc}
  \hline
  $\alpha$ & $l_{\mbox{opt}}$\\
  \hline
  1.125 & 0.0855 \\
  1.25 & 0.1874 \\
  1.5 & 0.3530 \\
  2 & 0.4666 \\
  3 & 0.4955 \\
  4 & 0.4991 \\
  5 & 0.4998 \\
  10 & 0.5000 \\ \hline
  \end{tabular}
  }
    \caption{Optimal choice of $l$ as a function of $\alpha$ (in Theorem \ref{ErrBounds}), from \protect \cite{HughettBounds}.}
  \end{center}
\end{table}

%% file: Tab2.tex
\begin{table}[t] \label{tab:Fourier_n}
\begin{center} {\scriptsize
  \begin{tabular}{ccc}
  \hline
  $\varepsilon$ & $N$ & time \\
  \hline
  $10^{-3}$ & 216 & 0.12 s\\
  $10^{-4}$ & 314 & 0.15 s\\
  $10^{-5}$ & 442 & 0.21 s\\
  $10^{-6}$ & 774 & 0.29 s\\
  \hline
  \end{tabular}
  }
    \caption{Normally distributed risk factors. The integer $N$ from Theorem \ref{ErrBounds} and the total time for the Fourier
    method computation of VaR$_\gamma$ as a function of $\varepsilon$, for $\gamma =
    0.01$.}
  \end{center}
\end{table}

%% file: Tab3.tex
\begin{table}[t] \label{tab:Fourier_t}
\begin{center} {\scriptsize
  \begin{tabular}{ccc}
  \hline
  $\varepsilon$ & $N$ & time \\
  \hline
  $10^{-3}$ & 1000 & 0.43 s\\
  $10^{-4}$ & 1508 & 0.65 s\\
  $10^{-5}$ & 2190 & 0.92 s\\
  $10^{-6}$ & 3070 & 1.28 s\\
  \hline
  \end{tabular}
  }
    \caption{Multivariate $t$-distributed risk factors. The integer $N$ from Theorem \ref{ErrBounds} and the total time for the Fourier
    method computation of VaR$_\gamma$ as a function of $\varepsilon$, for $\gamma =
    0.01$.}
  \end{center}
\end{table}

%% file: Tab4.tex
\begin{table}[t]
\begin{center}
  {\scriptsize
  \begin{tabular}{cccc}
  \hline
  $\varepsilon$ & $M$ &time (normal) & time (mult. $t$)\\
  \hline
  $10^{-3}$ & $6.5\cdot 10^{4}$ &1.76 s& 2.45 s\\
  $10^{-4}$ & $6.5\cdot 10^{6}$ &3 min & 4 min\\
  $10^{-5}$ & $6.5\cdot 10^{8}$ &4.5 h & 6.5 h\\
  \end{tabular}
  }
    \caption{The number of Monte Carlo samples $M$ needed to make $\mathbb{P}\left(|\gamma^*-\gamma|\leq \varepsilon\right) \geq 0.99$ as a function of $\varepsilon$. We also give the computation times for both the normal and the multivariate $t$-distributions.}

  \end{center} \label{tab:MC}
\end{table}